%% file: main_techreport.tex
\documentclass{sig-alternate-10pt}
\usepackage{url}
\usepackage{subfigure}
\usepackage{eurosym}

\newcommand{\eg}{{\it e.g.}}
\newcommand{\ie}{{\it i.e.}}

\newcommand{\bi}{\begin{itemize}}
\newcommand{\ei}{\end{itemize}}
\newcommand {\beq}{\begin{equation}}
\newcommand {\eeq}{\end{equation}}
\newcommand {\be}{\begin{enumerate}}
\newcommand {\ee}{\end{enumerate}}

\begin{document}

\title{Is Content Publishing in BitTorrent Altruistic or Profit-Driven?}

\numberofauthors{3}
\author{
\alignauthor
Ruben Cuevas\\
\affaddr{Univ. Carlos III de Madrid}\\
{\tt rcuevas@it.uc3m.es}
\alignauthor
Michal Kryczka\\
\affaddr{Institute IMDEA Networks and Univ. Carlos III de Madrid}\\
{\tt michal.kryczka@imdea.org}
\alignauthor
Angel Cuevas\\
\affaddr{Univ. Carlos III de Madrid}\\
{\tt acrumin@it.uc3m.es} \and
\alignauthor
Sebastian Kaune\\
\affaddr{TU Darmstadt}\\
{\tt kaune@kom.tu-darmstadt.de}
 \alignauthor
Carmen Guerrero\\
\affaddr{Univ. Carlos III de Madrid}\\
{\tt guerrero@it.uc3m.es}
\alignauthor
Reza Rejaie\\
\affaddr{University of Oregon}\\
{\tt reza@cs.uoregon.edu}
}

\maketitle

\begin{abstract}

BitTorrent is the most popular P2P content delivery application where individual users share various type of content with tens of thousands of other users. The growing popularity of BitTorrent is primarily due to the availability of valuable content without any cost for the consumers. However, apart from required resources, publishing (sharing) valuable (and often copyrighted) content has serious legal implications for user who publish the material (or publishers). This raises a question that whether (at least major) content publishers behave in an altruistic fashion or have other incentives such as financial. In this study, we identify the content publishers of more than 55k torrents in 2 major BitTorrent portals and examine their behavior. We demonstrate that a small fraction of publishers are responsible for 66\% of published content and 75\% of the downloads. Our investigations reveal that these major publishers respond to two different profiles. On one hand, antipiracy agencies and malicious publishers publish a large amount of fake files to protect copyrighted content and spread malware respectively. On the other hand, content publishing in BitTorrent is largely driven by companies with financial incentive. Therefore, if these companies lose their interest or are unable to publish content, BitTorrent traffic/portals may disappear or at least their associated traffic will significantly reduce.

\end{abstract}


\input{introduction}
\input{measurement_methodology}

\input{identity_major_publishers}

\input{signature_techreport}
\input{incentives}

\input{bussiness_model}
\input{application}
\input{related_work}
\input{conclusion}
\input{acknowledgement}

{\small
\bibliographystyle{plain}
\bibliography{rcuevas.bib}
}

\appendix
\input{seeding_time_estimation}

%
%
\end{document}

%% file: introduction.tex
\section{Introduction}

Peer to Peer (P2P) file-sharing applications, and more specifically
BitTorrent, are a clear example of \emph{killer} applications in the
last decade. For instance, BitTorrent is currently used by hundreds of
millions of users and is responsible for a large portion of the Internet traffic share \cite{CACHE_LOGIC}. This has attracted the attention of the research community that has mainly focused on understanding the technical aspects of BitTorrent functionality \cite{Guo05:BitTorrentMeasurement,Legout07:Clustering,kaune-2009} and improving its performance \cite{Piatek07:BitTyrant,Laoutaris2008:bitmax}. Moreover, some few papers have analyzed the demographics of BitTorrent \cite{ross_bt,biersackbittorrent,Pouwelse05:BitTorrent} and also the security  \cite{Sirivianos:free-riding} and privacy issues \cite{legout:spying, choffnes:communities}. However, the socio-economic aspects associated to BitTorrent in particular, and to other P2P file sharing systems in general, have received little attention despite of the importance that they have to the complete understanding of such kind of applications. This paper is a first step in this direction. 

The availability of free popular (often copyrighted) content that is
of interest to millions of people (e.g. recent TV Shows episodes,
Hollywood movies, etc) is the key pillar that makes BitTorrent an
extremely successful system. In this paper we study content publishing
in BitTorrent from a technical and more importantly socio-economic
point of view. In short, we try to unravel \emph{who} 
publishes content in BitTorrent, and \emph{why}.

For this purpose we rely on real data from a large scale measurement study
performed over two large BitTorrent Portals (Mininova and The Pirate
Bay). Our dataset ($>$300 GB) consits of information on more than 35M IP addresses and more than 55K published contents including the content publisher.

Using this dataset, we have first looked at the contribution of the different content publishers and conclude that just few publishers (around 100) are responsible of uploading 2/3 of the contents that serve 3/4 of the downloads in our major dataset. Furthermore, an important part of these major publishers consume few or even no contents, rather they dedicate their resources (almost) to only seed the published content. This is unusual behavior since standard BitTorrent users typically employ their resources for both seeding and downloading contents. 
Therefore, our observation reveals that major BitTorrent publishers present an anomalous behavior. This argument is reinforced after checking that many of the files published by these major publishers are copyrighted. Then, major publishers not only expend their resources without any apparent benefit, but they also face legal reactions due to the publication of copyrighted content
\cite{bt_arrest:1,bt_arrest:3}. These findings raise the following questions:  \emph{are these major publishers good citizenships that allocate a great deal of resources and assume legal risk for the good of the community?} or contrary, \emph{do they have any (still) unknown incentive to behave in this manner?}

To answer these questions we perform a systematic study of the aforementioned major publishers. We first discover the identity of these major publishers by looking at their associated usernames and IPs. This allows us to classify them into two different groups: \emph{fake publishers} publish a large number of fake content and \emph{top} publishers publish a large number of proper (often copyrighted) content.

Afterwards we study  main characteristics of these groups  such as the popularity of the content they publish and their seeding behavior (\ie~its signature). One one hand, our results reveal that the falseness of the content published by \emph{fake publishers} makes their swarms unpopular and obeys them to seed multiple torrents in parallel across long sessions. On the other hand, \emph{top publishers} are responsible of very popular contents for which they guarantee a proper seeding.

Finally we exploit the available information related to these publishers (\eg~in the BitTorrent portals) and conclude that \emph{fake} publishers are linked to antipiracy agencies
and malicious users, whereas half of \emph{top} publishers run their own web sites that report them economical benefits that are very substantial in some few cases.

In summary the main contributions of this paper are:

\bi
\item A simple measurement methodology to monitor the content
publishing activity in major BitTorrent portals. This methodology has
been used to implement a system that continuously monitors the content
publishing activity in The Pirate Bay portal. The data gathered is
made publicly available through a web interface.

\item The major portion of content publishing activity in BitTorrent
is concentrated in a relative small set of publishers (around 100) that are responsible of 2/3 of the published content and 3/4 of the downloads. This set of publishers can be further divided into three subsets that we name \emph{fake} publishers, \emph{altruistic top} publishers and \emph{profit-driven top} publishers. 

\item \emph{fake} publishers are set up by either anti-piracy agencies or malicious users and are responsible of 30\% of the content and 25\% of downloads. This means that these publishers sustain a continuous poisoning-like index attack \cite{ross:poisoning} against BitTorrent portals that based on our results affects to millions of downloaders.

\item \emph{profit-driven top} publishers own fairly profitable web sites. They use major BitTorrent portals such as The Pirate Bay as a platform to advertise their web site to millions of users. For this purpose they publish popular torrents where they attach the URL of their web site in various manners. The publishers linked to this business model are responsible of around 30\% of content and 40\% of downloads.
\ei


The rest of the paper is organized as follows. Section
\ref{sec:method} describes our measurement methodology. Sections
\ref{sec:major_publihsers} and \ref{sec:signature} are dedicated to
the identification of major publishers and  their main characteristics
(\ie~signature) respectively. In Section \ref{sec:incentives} we study
the incentives that major publishers have to perform this activity.
Section \ref{sec:bussiness} presents other players that also benefit
from content publishing. In Section \ref{sec:application} we describe our publicly available application to monitor content publishing activity in The Pirate Bay portal. Finally Section \ref{sec:rw} discusses related work and Section \ref{sec:conclusion} concludes the paper.

.

%% file: measurement_methodology.tex
\vspace{-0.5cm}
\section{Measurement Methodology}
\label{sec:method}
This section describes our methodology to identify the initial publisher of a file
that is distributed through a BitTorrent swarm. Towards this end, we first briefly 
describe the required background on how a user joins a BitTorrent swarm. 

\noindent
{\bf Background:}
A BitTorrent client takes the following steps to join the swarm associated with file
$X$. 
First, the client obtains the .torrent file associated to the desired swarm. 
The .torrent file contains contact information for the tracker that manages 
the swarm and the number of pieces of file $X$, etc.
Second, the client connects to the tracker and obtains the following information: 
$(i)$ the number of seeders and leechers that are currently connected to the swarm, 
and $(ii)$ N (typically 50) random IP addresses of participating peers in the swarm. 
Furthermore, if the number of neighbors is eventually lower than a given threshold 
(typically 20), the client contacts the tracker again to learn about other peers in the
swarm. 

To facilitate the bootstrapping process, the .torrent files are typically indexed at BitTorrent portals. Some of the major portals (\eg~The Pirate Bay or Mininova) index millions of .torrent files \cite{ross_bt}, classify them into different 
categories and provide a web page with detailed information (content category, 
publisher's username, file size, and file description). These portals also offer an RSS feed
to announce a newly published file. 
The RSS gives also some extra information such as content category, content size and username that published the .torrent file.

\noindent
{\bf Identifying Initial Publisher:}
In BitTorrent a content publisher is identified by its username \cite{ross_bt} and IP address.
The objective of our measurement study is to determine the identity of the initial 
publisher of a large number of torrents and to assess the popularity of each published file (\ie~the number and identity of peers who download the published file). 

Toward this end, we leverage the RSS feed to detect the availability of a new file on
major BitTorrent portals and retrieve the publisher's username. In order to obtain the publisher's IP address, we immediately download the .torrent file and connect to the associated tracker\footnote{Note that for most of the torrents we have used the Open BitTorrent tracker, that is the current major BitTorrent tracker.}. 
This implies that we often contact the tracker shortly after the birth of the associated 
swarm when the number of participating peers is likely to be small and the initial publisher
(\ie~seeder) is one of them. We retrieve the IP address of all participating peers as well
as the current number of seeders in the swarm. If there is only one seeder in the swarm
and the number of participating peers is not too large (\ie~$<$ 20), we obtain the bitfield of available pieces at individual peers to identify the seeder. Otherwise, reliably identifying the initial seeder is difficult because there are more than one seeder or the number of participating peers is large. 
We cannot directly contact the initial seeder that is behind a NAT box and thus unable
to identify the initial publisher's IP address in such cases.
Therefore, we are able to infer the publisher's username for all the crawled files. Furthermore, since we quickly detect the birth of a torrent through the RSS feed, we often 
(for 40\% of the files) contact the swarm when the number of participating peers are small and there is a single seeder that is not behind a NAT\footnote{Our investigations revealed two interesting scenarios for which we could not identify the initial publisher's IP address: 
{\em (i)} swarms that have a large number of peers shortly after they are added to the portal. We discovered that these swarms have already been published in other portals.
{\em (ii)} swarms for which the tracker did not report any seeder for a while or
did not report a seeder at all.}, thus being able to infer the publisher's IP address.

\begin{table}
\scriptsize
\hspace{-0.7cm}
\begin{tabular}[t]{l||lllll}
	       & Portal     & Start     & End       &\#Torrents &\#IP  \\
	       &            &           &          &            &addresses\\\hline\hline  
\emph{mn08}    & Mininova   & 09-Dec-08 & 16-Jan-09 &     -  /20.8K      & 8.2M    \\
\emph{pb09}    & Pirate Bay & 28-Nov-09 & 18-Dec-09 &   23.2K/10.4K      & 52.9K   \\
\emph{pb10}    & Pirate Bay & 06-Apr-10 & 05-May-10 &   38.4K/14.6K      & 27.3M   \\
\end{tabular}
\caption{Datasets Description}
\label{tab:measurements}
\vspace{-0.2cm}
\end{table}

Therefore, after the first connection to the tracker, we can identify the identity of
the initial publisher (username and/or IP address) in most cases and determine basic properties of the published
content.
At this point, we periodically query the tracker in order to obtain the IP addresses of the participants in the content swarm and always solicit the maximum number of IP addresses (\ie~200) from the tracker. To avoid being blacklisted by the tracker, we issue our query at the maximum rate that is allowed by the tracker (\ie~1 query every 10 to 15 minutes depending on the tracker load).  Given this constraint, we query the tracker from several geographically-distributed 
machines so that the aggregated information by all these machines provides an adequately high resolution view of participating peers and their evolution over time. We continue to monitor a target swarm until we receive 10 consecutive empty replies from the tracker.
We use MaxMind Database \cite{maxmind:page} to map all the IP addresses (for both publishers and downloaders) to their corresponding ISPs and geographical location.

\subsection{Dataset}
We used the previously described methodology to identify a large number of BitTorrent
swarms at two major BitTorrent portals, namely Mininova and The Pirate Bay. Each one of
these portals was the most popular BitTorrent portal at the time of of our measurement
according to Alexa ranking.
Table \ref{tab:measurements} shows the main features of our three datasets 
(1 from Mininova and 2 from the The Pirate Bay) including
the start and end dates of our measurement, the number of torrents for which we identified
the initial publisher (username/IP address), and the total number of discovered IP addresses associated for all 
the monitored swarms. We refer to these datasets as \emph{mn08}, \emph{pb09} and
\emph{pb10} throughout this paper. We note that dataset \emph{mn08} does not contain
the username of initial publishers, whereas \emph{pb09} queries the tracker just once after we identify the file through the RSS.

%% file: identity_major_publishers.tex
\section{Identifying Major Publisher}
\label{sec:major_publihsers}

As stated before a publisher can be identified by its username and/or IP address. In our analysis, we identify individual publishers by their username since the username is expected to remain consistent across different torrents, although we will make some exceptions (See \emph{fake} publishers below in this Section). Furthermore, in case of \emph{mn08} we use the IP address since we lack of username information.

\begin{table*}

\tiny

\begin{tabular}[t]{lll|lll|lll}

mn08& &  & pb09 & & & pb10 & &  \\\hline\hline
ISP  & type & \% & ISP & Type  & \% & ISP & Type &\% 		 \\\hline
OVH	 & Hosting Provider & 13.31 & OVH & Hosting Provider & 24.76 & OVH & Hosting Provider & 15.16 \\
Comcast & Commercial ISP & 4.69 & Comcast & Commercial ISP & 3.67 & SoftLayer Tech.& Hosting Provider & 4.52\\
Keyweb & Hosting Provider& 3.18 & Road Runner & Commercial ISP & 2.3 & FDCservers & Hosting Provider & 3.64\\
Road Runner & Commercial ISP & 3.03 & Romania DS & Commercial ISP & 2.27 &  Open Computer Network & Commercial ISP & 3.59\\
NetDirect & Hosting Provider & 2.44 &  MTT Network  & Commercial ISP & 1.95 & tzulo & Hosting Provider & 3.36\\
Virgin Media & Commercial ISP & 2.42 & Verizon & Commercial ISP & 1.64 & Comcast & Commercial ISP & 2.86\\
NetWork Operations Center & Hosting Provider & 2.39 & Virgin Media & Commercial ISP & 1.49 & Cosema & Commercial ISP & 2.25\\
SBC & Commercial ISP & 2.38 & SBC & Commercial ISP & 1.41 & Telefonica & Commercial ISP & 2.22\\
Comcor-TV & Commercial ISP & 2.33 & NIB & Commercial ISP & 1.26 & Jazz Telecom. & Commercial ISP & 2.07 \\
Telecom Italia & Commercial ISP & 2.02 & tzulo & Hosting Provider & 1.14 & 4RWEB & Hosting Provider & 2.06 \\
\end{tabular}
\caption{Content Publishers Distribution per ISP}
\label{tab:publisher_per_ISP}
\end{table*}

\begin{figure}[t]
\centering
\includegraphics[width=2.8in]{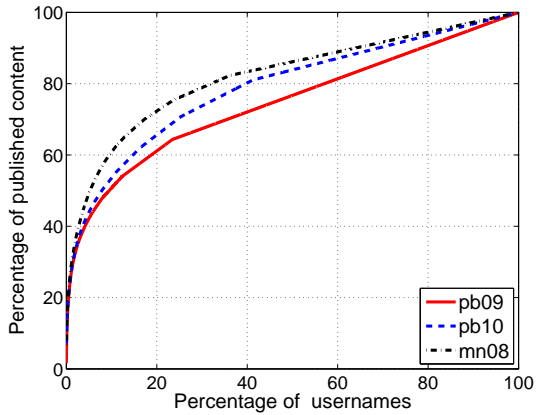} 
\caption{Percentage of content published by the top x\% publishers}
\label{fig:fed_content}
\vspace{-0.3cm}
\end{figure}

\subsection{Skewness of Contribution}

First, we examine the level of contribution (\ie~the number of published files) across identified content publishers in each dataset.
Figure \ref{fig:fed_content} depicts the percentage of files that are published by the top x\% of publishers. 
We observe that the top 3\% of BitTorrent publishers contribute roughly 40\% of published content. Furthermore, if we carefully look at the IP addresses of the top-100 (\ie~3\%) publishers in our \emph{pb10} dataset, we realize that an important part of them either do not download any content (40\%) or download less than 5 files (80\%).
This large contribution of resources (bandwidth, storage, or content) among top BitTorrent publishers does not seem to be justified by peers' altruism for two simple reasons:
\vspace{0.15cm}

\noindent \textbf{{\em - Required Resources/Cost}}:
publishing a large number of content requires a significant amount of processing and bandwidth. For example, some of these major content publishers  (\eg~\url{www.eztv.it}) recommend to allocate at least 10Mbps in order to sustain the seeding of few (around 5) files in parallel.
\vspace{0.1cm}

\noindent \textbf{{\em - Legal Implications}}: 
As other studies have reported \cite{legout:spying} and we confirm in our datasets, 
a large fraction of content published by major publishers is  copyrighted
material (recent movies or TV series). Thus, publishing these files is likely to
have serious legal consequences for these publishers \cite{bt_arrest:1,bt_arrest:3}.
\vspace{0.15cm}

This raises the question that \emph{why these small fraction of entities allocate a great deal of (costly) resources to contribute many files into BitTorrent sessions despite potential legal implications?} 
\emph{Is this level of contribution simply derived by an altruistic behavior of good
citizens or there are other incentives?}. We answer these questions in Section \ref{sec:incentives}.

\subsection{Publishers' ISPs}

To help identify content publishers in our dataset,
we determine the ISP that host each major publisher and use that
information to assess the type of service (and available resources)
that a publisher is likely to have.
Toward this end, we map the IP address for a publisher 
in each dataset to its corresponding ISP using the MaxMind database \cite{maxmind:page}.
We then examine publicly available information about each ISP (\eg~its web
page) to determine whether it is a commercial ISP or a hosting provider.
We performed these analysis only for the top-100 (roughly 3\%) of publishers
since they are mostly of interest and the collection of the required 
information is a tedious task.

We observe that 42\% of top-100 publishers in \emph{pb10} (35\% 
of top-100 in \emph{pb09}) are located in ten hosting services.
Furthermore, more than half of these publishers are concentrated at
a particular hosting service (OVH), namely 22\% of top-100 in \emph{pb10} 
and 20\% of top-100 in \emph{pb09}. Since we do not have publishers'
username for \emph{mn08}, we examine top-100 publishers based on their
IP addresses (\ie~can not assess aggregated contribution of a user
with different IP addresses). 77\% of top-100 publishers in \emph{mn08}
are also located in hosting services and 45\% of them are specifically
at OVH.

In short, our analysis reveals that a significant fraction of major
publishers are located at a few hosting services and a large percentage 
of them at OVH.

We also examine the contribution of BitTorrent publishers at the
ISP-level by mapping all the publishers to their ISPs and identify 
the top-10 ISPs based on their aggregate published content
for each dataset as shown in Table \ref{tab:publisher_per_ISP}.
This table reveals that content publishers who are located at 
a particular hosting provider, namely OVH,
have consistently contributed a significant fraction of published 
content at major BitTorrent portals. There are also several 
commercial ISPs (\eg~Comcast) in Table \ref{tab:publisher_per_ISP} 
with a much smaller contribution.

To assess the difference between users from hosting providers
and commercial ISPs, we compare and contrast all publishers that
are located at OVH and Comcast as a representative ISP for each
class of publishers in Table \ref{tab:feeders}.
This table demonstrates the following two important differences:
First, the aggregate contribution of each publisher at OVH is on average
a few times larger than Comcast publishers. 
Second, Comcast publishers are sparsely scattered across many /16 IP prefixes 
and many geographical locations in the US whereas OVH publishers are concentrated 
in a few /16 IP prefixes and a handful of different locations in Europe (the location 
of OVH's data centers). In essence, the published content by Comcast comes from a large
number of typical altruistic users where each one publishes a small number of 
items from their home or work. In contrast, OVH publishers appear to be paying 
for a well provisioned service to be able to publish a much larger number of files.
We have also examined consumer peers in captured torrents and did not observe
the presence of OVH users among the consuming peers. 

{\em In summary, the examination of ISPs that host major BitTorrent publishers suggests that these publishers are located either at a few hosting providers (with a large concentration at OVH) or at commercial ISPs.
These publishers contribute a significantly larger number of files than average publishers. Furthermore, publishers who are located at hosting providers do not consume published content by other publishers.}

\begin{table}

\small

\centering



\begin{tabular}[tb]{l||lllll}
            & Fed         & IP addr   & /16 IP  & Geo  \\ 
	    & torrents	      &	          &	Pref.	& Loc.	     \\\hline\hline
OVH (mn08)   & 2766      & 164      & 5           & 2       \\ 
Comcast (mn08) & 976     & 675      & 269         & 400     \\\hline 
OVH (pb09)   & 2577      & 78       & 5           & 2       \\ 
Comcast (pb09) & 382   & 198      & 143         & 129      \\\hline
OVH (pb10)   & 2213    & 92       & 7           & 4       \\ 
Comcast (pb10) & 408   & 185      & 139         & 147      \\
\end{tabular}
\caption{Characteristics of OVH feeders and Comcast feeders in \emph{mn08} and \emph{pb09} datasets.}
\vskip -2mm
\label{tab:feeders}

\end{table}

\subsection{Fine Grained Look at Major Publishers}

Now, we examine the mapping between username and IP address of top-100
content publishers in the \emph{pb10} dataset in order to gain some
insight about major publisher behavior, and we discover the following interesting points: 

First, if we focus on the top-100 IP addresses who have published the largest number  of files, only 55\% of them are used by a unique username. The remaining 45\% of IP addresses of major publishers are mapped to a large number of usernames.
We have carefully investigated this set of IP addresses and discovered that they use either hacked or manually  created (\ie~random name) username accounts to inject fake content.  These publishers appear to be associated with anti-piracy agencies or malicious users. The former group tries to avoid the distribution of copyrighted content whereas the latter attempts to disseminate malware. We refer to these publishers as {\em fake publishers}. Surprisingly, fake publishers are responsible of 1030 usernames (around 25\%), 30\% of the content and 25\% of the downloads in our \emph{pb10} dataset. Then, major BitTorrent portals are suffering from a systematic poisoning index attack \cite{ross:poisoning} that affects to more than 1/4 of the published content. The portals fight this phenomenon by removing the fake contents as well as the user accounts used to publish them\footnote{we exploit this fact to identify if a username has been used by a fake publisher.}. However, this technique does not seem to be enough effective since millions of users initiate the download of fake content.
Finally, it is worth noting that most of the fake publishers perform their activity from three specific hosting providers named \emph{tzulo}, \emph{FDC Servers} and \emph{4RWEB}.
Due to the relevant activity of these publishers we study them as an independent group in the rest of the paper.

Second, the inspection of the top-100 usernames who publish the largest number of files shows that only 25\% of them operate from a single IP. The remaining 75\% of top usernames utilize multiple IPs  and can be classified into the following common cases: {\em (i)} 34\% of usernames with multiple IP addresses use a reduced number of IPs (5.7 in average) from hosting providers in order to provide the required resources for seeding a large number of files.$(ii)$ 24\% of usernames with multiple IP addresses (13.8 in average) located in a single commercial ISP. Their mapping to multiple IP addresses must be due to the periodical change of their assigned IP address by their ISP. $(iii)$ The other 16\% of these usernames are mapped to multiple IP addresses (7.7 in average) at different commercial ISPs. These are users who inject content from various locations (\eg~home and work computer).
To minimize the impact of abnormal publishers, we removed the 16 usernames from the top-100 usernames that appeared to be compromised and used for publishing fake content (and will be analyzed separately). We refer to the resulting group as \emph{Top}. Note that the \emph{Top} publishers are responsible of 37\% of the content and 50\% of the total downloads in our \emph{pb10} dataset.

In summary, the major portion of the content comes from two reduced group of publishers: \emph{Top} publishers and \emph{Fake} publishers that collectively are responsible of 2/3 of the published content and 3/4 of the downloads. In the rest of the paper we devote or effort to characterize these two groups.

%% file: signature_techreport.tex
\section{Signature of Major Publisher}
\label{sec:signature}
Before we investigate the incentives of major BitTorrent publishers, we 
examine whether they exhibit any other distinguishing features,
\ie~whether major publishers have a distinguishing signature. 
Any such distinguishing features could shed some light on the underlying 
incentives of these publishers. Toward this end, in the next few subsections, we examine the following 
characteristics of major publishers in our datasets: 
{\em (i)} the type of published content,
{\em (ii)} the popularity of published content, and 
{\em (iii)} the availability and seeding behavior of a publisher.

To identify distinguishing features, we examine the above characteristics across
the following three {\em target groups} in each dataset: 
all publishers (labeled as ``All"),
all fake publishers (labeled as ``Fake") and
all top-100 (non-fake) publishers  regardless of their ISPs (labeled as ``Top"). We also examine the break down of top publishers based on their ISPs
into hosting providers and commercial ISPs, labeled as ``Top-HP" and ``Top-CI",
respectively\footnote{All the top-** results for dataset {\em mn08} are determined
based on IP address since we do not have username information for publishers
in this dataset as we stated earlier. Also without username information for
each publisher in {\em mn08} dataset, we can not identify fake publishers.}. 

\begin{figure}[t]

\subfigure[\emph{mn08}]{\hspace{-0.70cm} \includegraphics[width=1.70in]{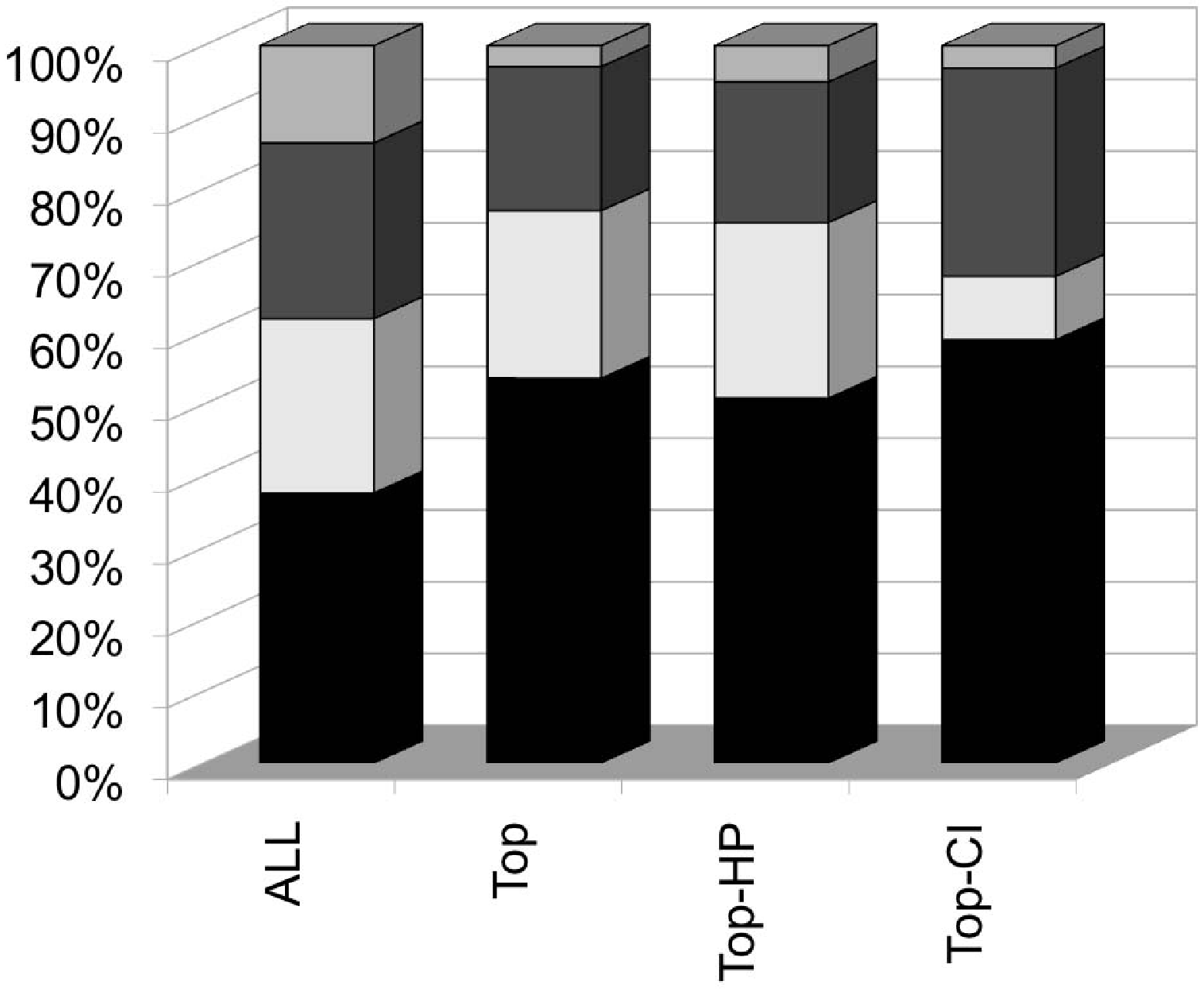} \label{subfig:bar_m08}}
\hspace{-0.50cm}
\subfigure[\emph{pb10}]{\includegraphics[width=2in]{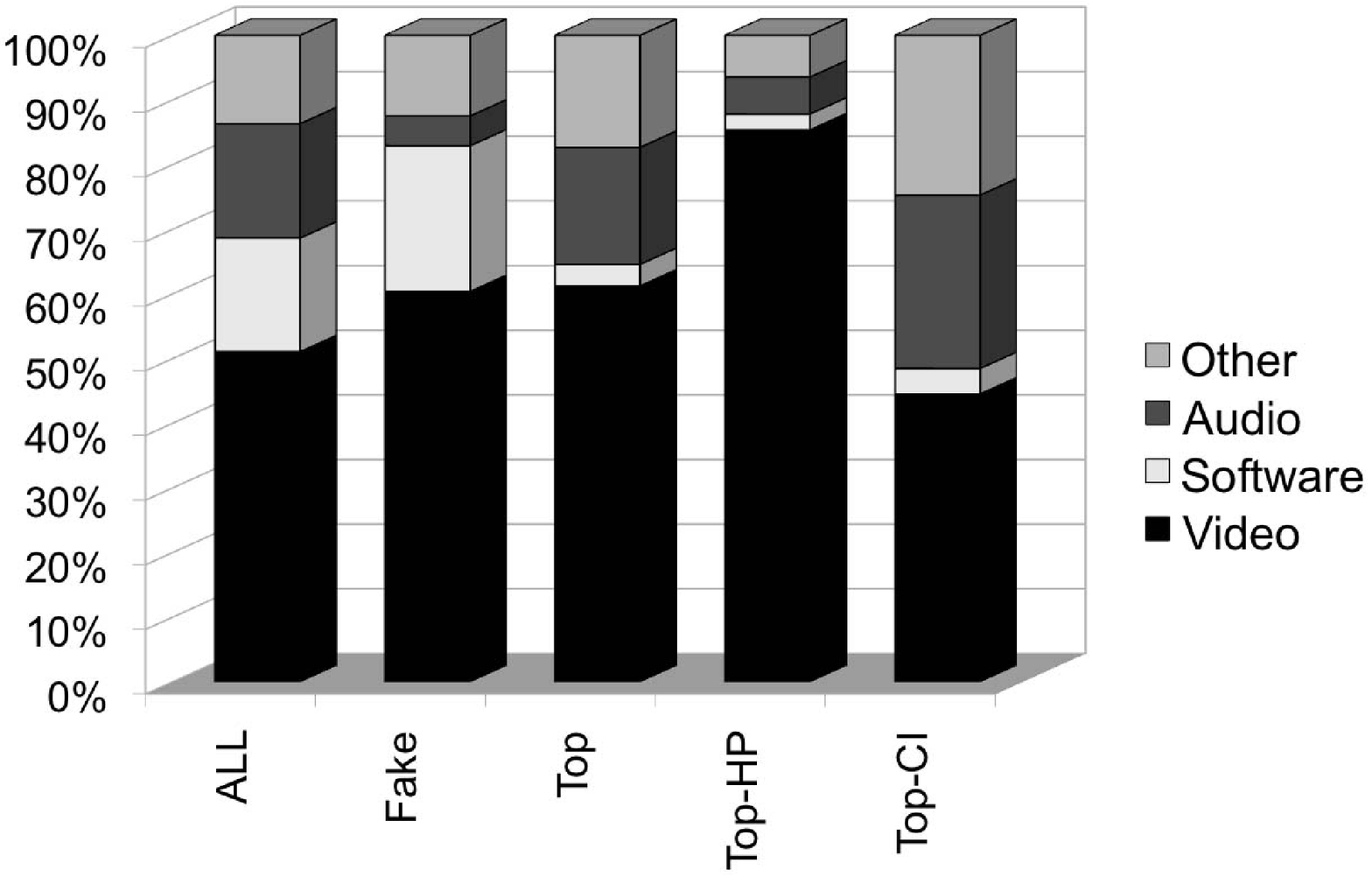} \label{subfig:bar_pb10}}
\caption{Type of Content Published distribution for the different set of publishers: \emph{all}, \emph{fake}, \emph{top}, \emph{top-hp} and \emph{top-ci}}
\label{fig:type_of_content}
\vspace{-0.3cm}
\end{figure}

\subsection{Content Type}
We leverage the reported content type by each publisher to classify the published content
across different groups. Figure \ref{fig:type_of_content} depicts the break down of published
contents across different type for all publishers in each target group for our Mininova and our major Pirate Bay datasets. This figure reveals a few interesting trends as follows:

First, Video (composed mainly by Movies, TV-Shows and Porn content) constitutes a significant fraction of published files across most groups with some important differences. The percentage of published video across
all publishers is around 37\%-51\% but it is slightly larger among top publishers.
However, video is clearly a larger fraction of published content by the top publishers located in hosting providers in our \emph{pb10} dataset.
Fake publishers primarily focus on Videos (recent movies and shows) and Software content. This supports our earlier observation that these publishers consist of antipiracy agencies and
malicious users where the former group publishes a fake version of e.g. last movies and the latter provides
software that contains malware.

\subsection{Content Popularity}
\label{subsec:popularity}
\begin{figure}[t]
\centering
\includegraphics[width=3.5in,height=2in]{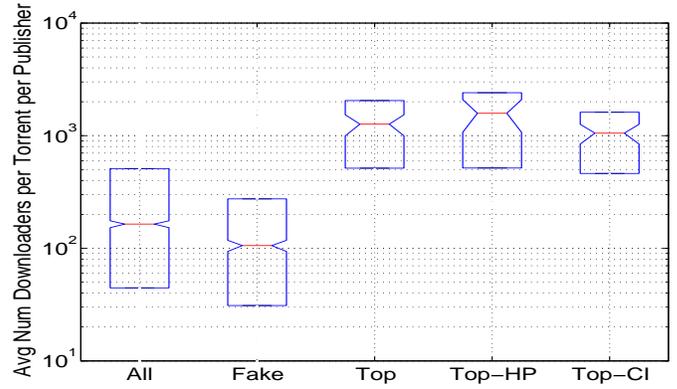}  
\caption{Avg Num of Downloaders per torrent per publisher for \emph{all}, \emph{fake}, \emph{top}, \emph{top-hp}, \emph{top-ci}}
\label{fig:avg_num_downloaders}
\vspace{-0.3cm}
\end{figure}

The number of published files by a publisher shows only one dimension of its contribution
to BitTorrent. The other equally important issue is the popularity of each published
content (\ie~the number of downloaders regardless of their download progress) by individual publishers. 
Figure \ref{fig:avg_num_downloaders} shows the box plot of the distribution of average downloaders per torrent per publisher across all publishers in each target group where each box presents 25th, 50th and 75th percentiles. 

On the one hand, the median popularity of top publishers' torrents is 7 times higher than standard users (represented by \emph{All}). If we look more closely to the top publishers, those located in hosting providers publish torrents in median 1.5 times more popular than those located in commercial ISPs. Then, those top publishers with more resources publish typically more popular contents.
On the other hand, fake publishers' torrents are the most unpopular among the studied groups. This is because the portals actively monitor the torrents and immediately remove the content identified as fake, thus avoiding new users downloading it. Furthermore, users quickly realize the fake nature of these torrents and report this info on blogs that also prevent other users from downloading them.

In summary, top publishers are responsible of a larger number of popular torrents. This has a multiplicative effect making top publishers that inject 37.5\% of the content being responsible of a higher percentage of downloaders (around 50\%). The low popularity of fake publishers' torrents produces a contrary effect: they are responsible of 30\% of the content but only 25\% of the downloads.

\begin{figure*}[t]
\centering
\subfigure[\emph{Average Seeding Time per Content per Publisher}]{\includegraphics[width=2.2in]{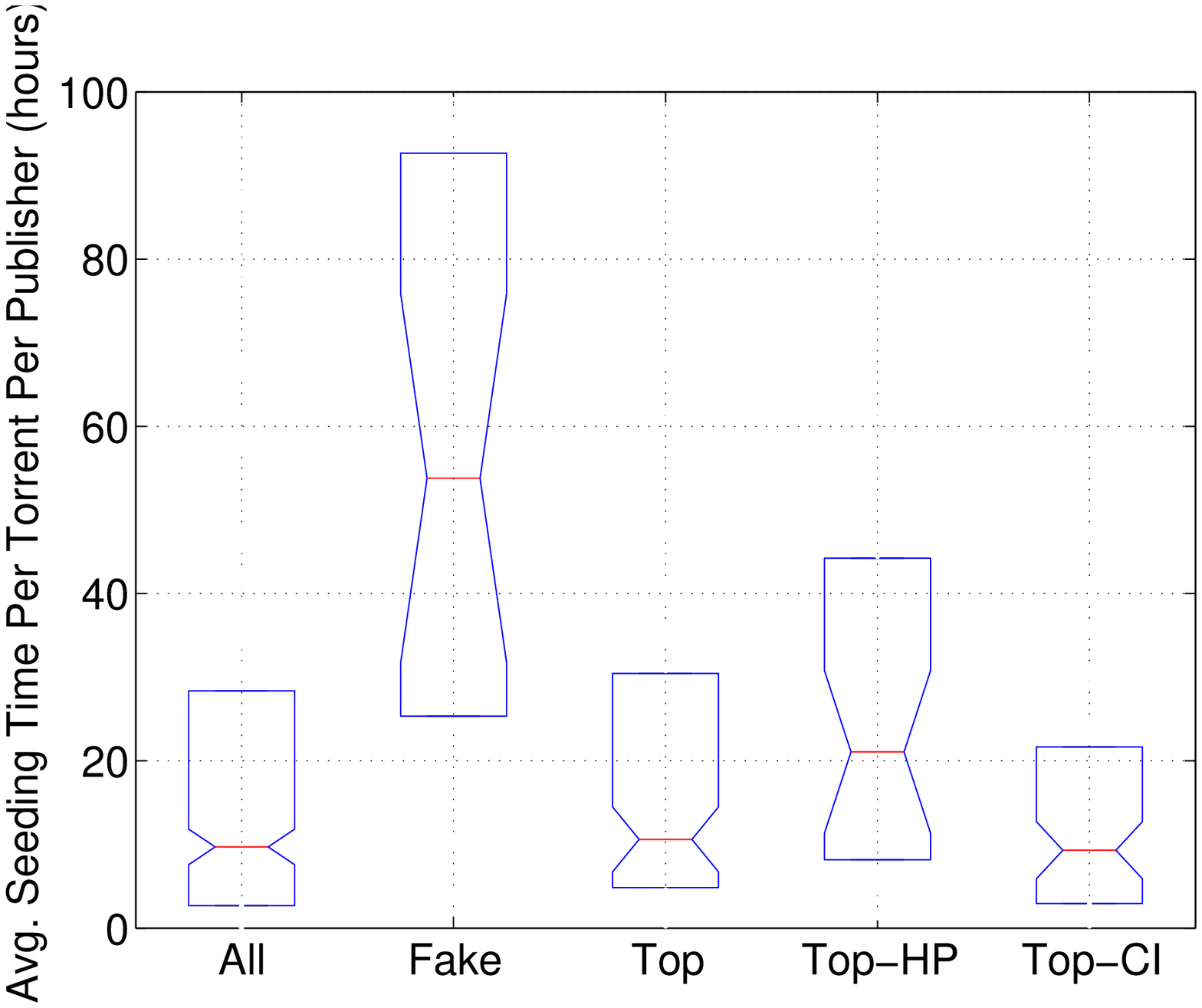}\label{fig:seeding_time}}
\subfigure[Average Number of Parallel Seed Torrents Per Publisher]{\includegraphics[width=2.2in]{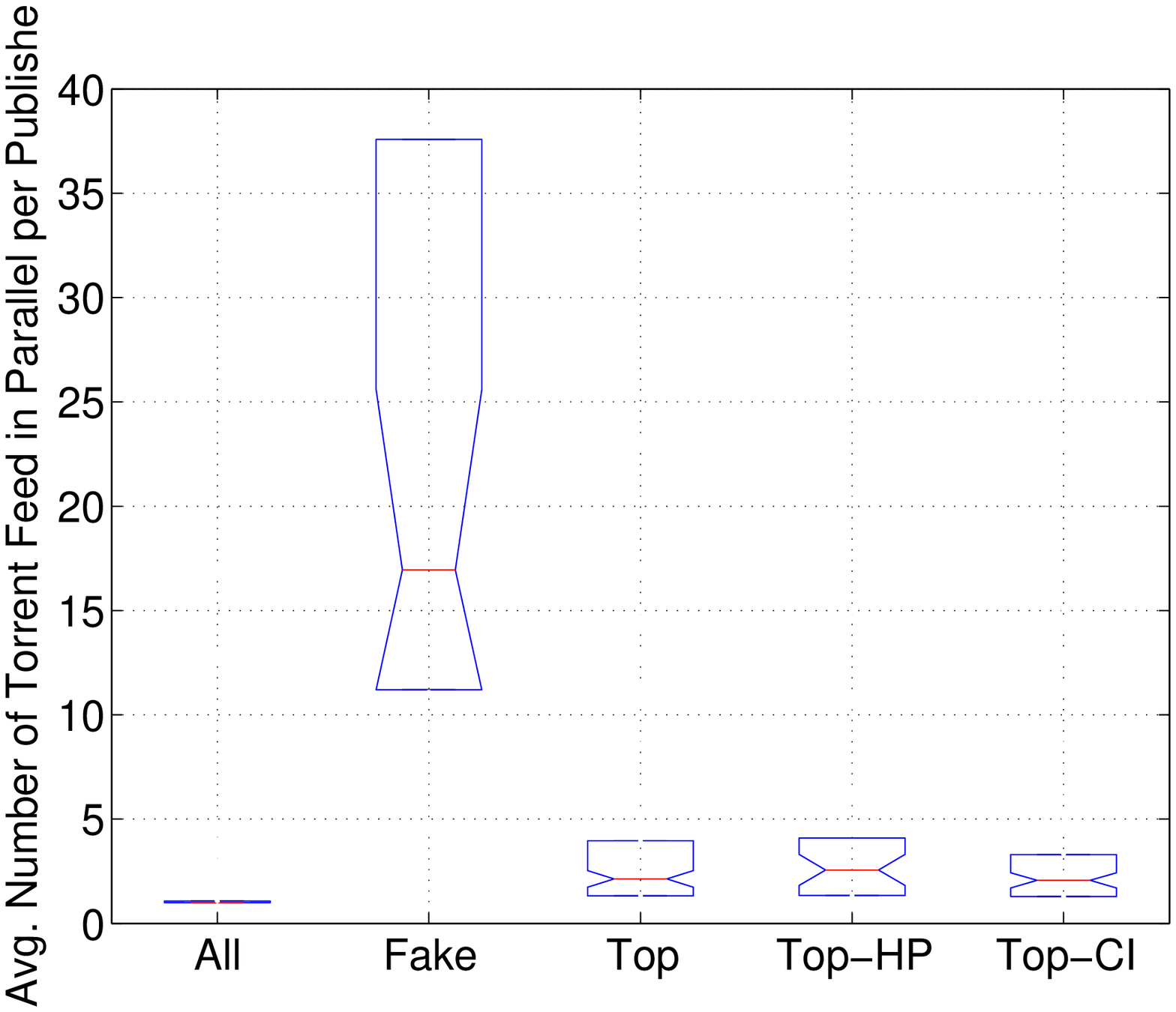}\label{fig:parallel}}
\subfigure[Average Global Session Time Per Publisher]{\includegraphics[width=2.2in]{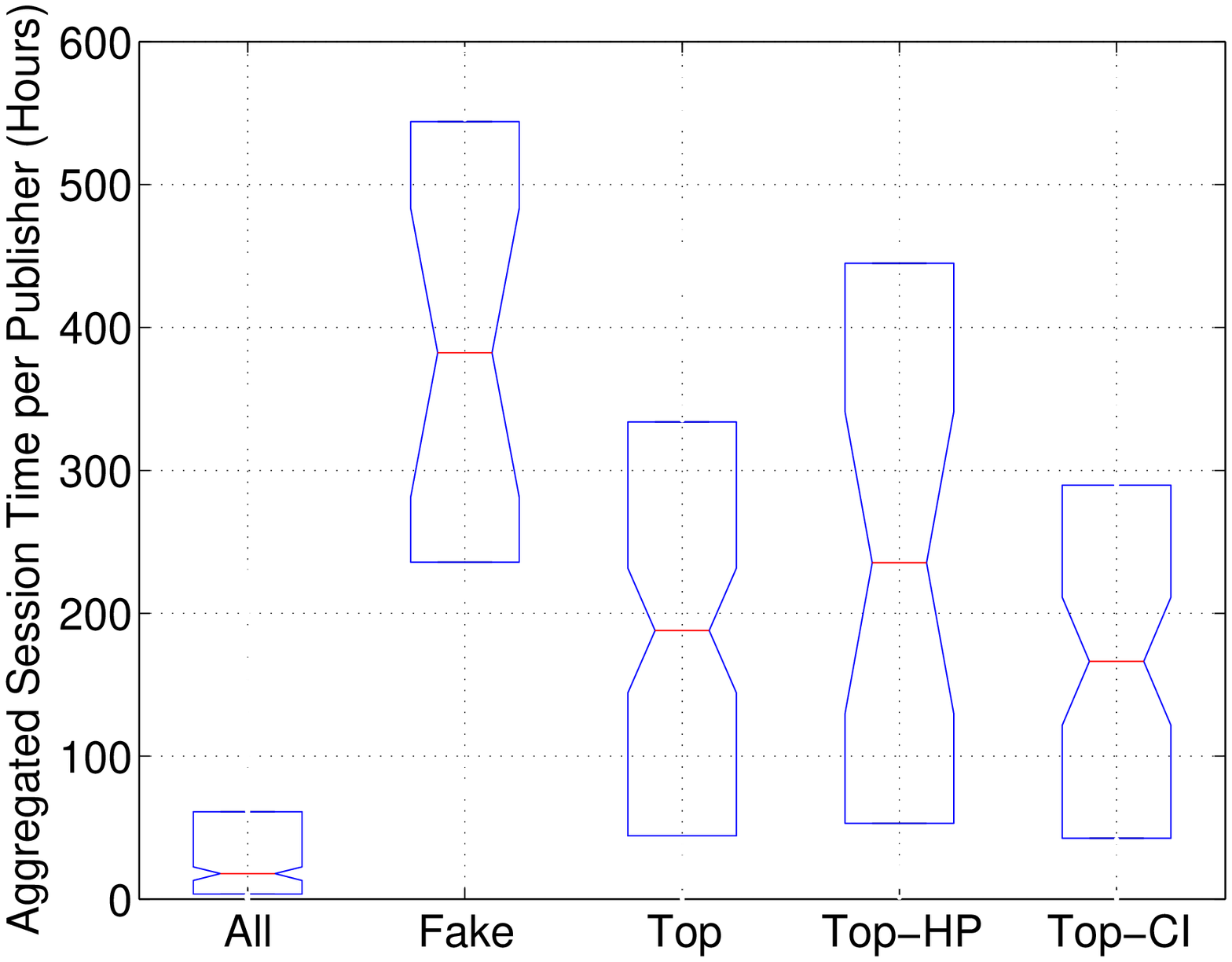}\label{fig:session}}
\caption{Seeding Behaviour for \emph{all}, \emph{fake}, \emph{top}, \emph{top-hp} and \emph{top-ci}}
\label{fig:seeding_behaviour}
\end{figure*}

\subsection{Seeding Behavior}
\label{subsec:behaviour}
We characterize the seeding behavior of individual BitTorrent publishers
in our target groups using the following metrics: 
{\em (i)} Average seeding time of a publisher for its published content, $(ii)$ Average number of parallel seeded torrents and $(iii)$ aggregated session time of a publisher across all its torrents.
Since calculating these properties requires detailed analysis of our dataset that are
computationally intensive, we are unable to derive these values for all publishers, instead we have used a random set of 400 publishers to represent the standard behaviour of the publishers\footnote{For consistency and readability purposes we use the label ``All'' to refer to this random subset of all publishers.}.

In order to compute these metrics we need to estimate the time that a specific publisher has been connected to a torrent (in one or multiple sessions). Since each query to the tracker just reports (at most) a random subset of 200 IPs, in big torrents ($>$200 peers), we need to perform multiple queries in order to assess the presence of the publisher in the torrent. In Appendix \ref{sec:seeding_time} we detail the technique used to compute the session time of a specific user in a specific torrent.


\noindent
{\bf Average Seeding Time:}
We measure the duration of time that a publisher stays in a torrent since its birth  to seed the content. In general, a publisher can leave the torrent once there is an adequate fraction of other seeds. 
Figure \ref{fig:seeding_time} depicts the summary distribution of average seeding time
across all publishers in each target group. This figure demonstrates the following points:
First, the seeding time for fake publishers is significantly longer than 
publishers in other groups. Since these publishers do not provide the actual
content, the initial fake publisher remains the only seed in the session (\ie~ other users do not help in seeding fake content) to keep the torrent alive.
Second, focusing on top publishers the Figure shows that they typically seed a content for a few 
hours. However, the seeding time for top publishers from hosting providers is clearly
longer than top publishers from commercial ISPs.  This suggests that publishers at
hosting providers are more concerned about the availability of their published content.

\noindent{\bf Average number of Parallel Torrents:}
Figure \ref{fig:parallel} depicts the summary distribution of the average number of torrents
that a publisher seeds in parallel across publishers in each target group.
This figure indicates that fake publishers seed many torrents in parallel. We have seen that fake publishers typically publish a large number of torrents, furthermore no one helps them on seeding task. Thus if they want to keep alive all these torrents they have to seed all of them in parallel.

The result for top publisher shows that the typical number of seeded torrents in
parallel is the same (around 3 torrents) for top publishers regardless of their
location. However, we expect that a regular publisher  seed concurrently only 1 file.

\noindent{\bf Aggregated Session Time:}
We have also quantified the availability of individual publishers by estimating
the aggregated session time that each publisher is present in the system  across all published torrent. 
Figure \ref{fig:session} shows the distribution of this availability measure across
publishers in each target group. As expected fake publishers present the longest aggregated session time due to their obligation to continuously seed their contents to keep them alive.
If we focus on top publishers, they exhibit a typical aggregated session 10 times longer than standard users. Furthermore, top publishers at hosting services are clearly more available than those from commercial ISPs.


\subsection{Summary}

BitTorrent content publishers can be divided into normal/altruistic users that are average users playing the role of publisher and consumer in BitTorrent. Whereas fake and top publishers publish a significant amount of files. One one hand, fake publishers publish a large number of unpopular Video and Software contents from a reduced number of hosting providers. Due to the falseness of their files, they need to seed multiple files across long sessions to keep these alive. We have stated that this behaviour responds to the interest of antipiracy agencies and malicious users that set up the fake publishers. We validate it in the next section.

On the other hand, top publishers often focus on video and they are located in either hosting facilities or commercial ISPs. Furthermore their content is popular, they stay long session in the system and ensure proper seeding of the content. All these evidences suggest that these top publishers are interested in visibility of the posted content (in attracting a large number of users). The behavior and incurred cost by these participants cannot be considered as altruistic especially since they typically do not consume but just inject content into the system.  Therefore, the  most likely conceivable motivation for the behaviour of these top publishers is financial incentive.  We examine this hypothesis in the rest of the paper.

%% file: incentives.tex
\vspace{-0.05cm}
\section{Incentives of Major Publishers}
\label{sec:incentives}

We have first analyzed fake publishers. We first look at the name of
the published files that are typically catchy titles (\eg~ recent
released Hollywood movies). Furthermore, we have downloaded a few of
their files. We note that in most of the cases the file had been
already removed\footnote{We tried to download the file after the
measurement study was performed and we had made a first analysis of
the data. This typically occurred a few weeks since the publication of
the content. Therefore, in most of the cases the content was anymore
no longer.}. The few downloaded files were indeed fake contents. Some
of them were modified with the inclusion of anti-piracy
advertisement/messages whereas some others led to malware
software\footnote{The content was a video that pointed to an specific
software to be played. This software was a malware
(\eg~\url{http://flvdirect.com/}.}. These observations validate our
assessment that behind fake publishers are anti-piracy agents and
malicious users. On one hand, the anti-piracy agents publish fake
files with the name of the copyrighted contents they want to protect.
On the other hand, the malicious users publish content with catchy
titles since these have the ability of attracting a larger population
of potential victims. Therefore, we have clearly characterized the
incentives of fake publishers. In the rest of this section we
characterize the major (non-fake) publishers.

Our goal is to identify the incentives of major (non-fake) BitTorrent publishers
to commit a significant amount of resources and deal with likely legal
implications of publishing copyrighted content. We believe that the
behavior of these publishers may not be altruistic and may be driven
by a financial interest. More specifically, our hypothesis is that
these publishers leverage major BitTorrent portals as a venue to
freely attract downloading users to their web sites.
To verify this hypothesis, we conduct an investigation to gather 
the following information about each one of the \emph{top} (\ie~top-100 non-faked) publishers:
\vspace{0.1cm}

\noindent{\em - Promoting URL}: the URL that downloaders of a published content may encounter,

\noindent{\em - Publisher's Username}: any publicly available information about
the username that a major publisher uses in The Pirate Bay portal, and

\noindent{\em - Business Profile}: offered services (and choices) at the promoting URL.
\vspace{0.1cm}

\noindent
{\bf Promoting URL:} 
We emulate the experience of a user for downloading a 
few randomly-selected files published by each top publisher to determine 
whether and where they may encounter a promoting URL. We identified three places where
publishers may embed a promoting URL:
{\em (i)} name of the downloaded file (\eg~user mois20 name all his files as \emph{filename-divxatope.com}, thus advertising the url \url{www.divxatope.com}),
{\em (ii)} the textbox in the web page associated with each published content,
{\em (iii)} name of a text file that is distributed with the actual content
and is displayed by the BitTorrent software when opening the .torrent file.

Our investigation indicates that the second approach (using the textbox) is
the most common technique among the publishers.
 
\noindent
{\bf Publisher's Username:}
We browsed the Internet to learn more information about the username associated
with each top publisher. First, the username is in some cases directly related to the URL (\eg user UltraTorrents whose url is \url{www.ultratorrents.com}). This exercise also reveals whether this username  publishes on other major BitTorrent portals in addition to The Pirate Bay. Finally, posted information in  various forums could reveal (among other things) the promoting web site.

\noindent
{\bf Business Services:} 
We characterize the type of services offered at the promoting URL and ways 
that the web site may generate income (\eg~posting ads).
We also capture the exchanged http headers between a web browser and the 
promoting URL to identify established connections to third-party
web sites (\eg~redirection to ads web sites or some third party aggregator) using the technique described in \cite{Krishnamurthy:headers}.

\subsection{Classifying Publishers}
Using the methodology described, we examined a few published torrents for each one of the
top publishers as well as sample torrents for 100 randomly selected 
publishers that are not in the top-100, called {\em regular publishers}.
On the one hand, we did not discover any interesting or unusual behavior in torrents
published by regular publishers and thus conclude that they behave in an altruistic manner. 
On the other hand, a large fraction of seeded torrents by the top publishers 
systematically promote one or more web sites for financial incentives. 
Our examination revealed that these publishers often include a
promotional URL
in the textbox of the content web page and some of them use other of the described techniques as well.
We classify these top publishers based on their business profile (\ie~type of business they run based
on the content of the promoting web sites) and describe how they leverage 
BitTorrent portals to intercept and redirect users to their web sites.
 


\noindent 
\textbf{Private BitTorrent Portals/Trackers:}
A subset of major publishers, 26\% of top, own their BitTorrent portals that are in some cases associated with private trackers \cite{powelse:private_tracker}. These private trackers guarantee a better user experience in terms of download rate (compared to major open BitTorrent Portals) but require clients to maintain
certain seeding ratio. More specifically, each participating BitTorrent client is required 
to seed content proportional to the amount of data they download across multiple torrents.
To achieve this goal, users are required to register in the website and login
before downloading the torrent files. 
 
The publishers in this class publish 18\% of all the content and more impressively they are responsible of 29\% of the downloads. This means that around 20 publishers are responsible of almost 1/3 of the downloads. 

2/3 of these publishers advertise the URL in the textbox at the content web page, whereas 1/4 use some of the other described techniques (note that some of them use multiple techniques to advertise their URL).

Furthermore, they appear to obtain financial profit in three different ways: 
$(i)$ Posting advertisement in their website, $(ii)$ Seeking donations from visitors to
continue their basic service, and $(iii)$ Collecting a fee for VIP access that allows the client to download any content without sustaining any kind of seeding ratio.

These publishers typically inject video, audio and application content into BitTorrent
portals. 
Interestingly, a significant fraction of publishers in this class
(40\%) publish content in a specific language (Italian, Dutch, Spanish or Swedish) 
and specifically a 66\% of this group are dedicated to Spanish content. This finding is consistent with
prior studies that reported a high level of copyright infringement in Spain  \cite{baytsp:piracy_report}.

\noindent 
{\bf Other Web Sites}: Another class of top publishers (24\%) promote URLs that are associated with web sites that are hosting images (\eg~\url{www.pixsor.com}), forums or even religious groups (\eg~\url{lightmiddleway.com}). These publishers inject 8\% of the content in the system and are responsible of 11\% of the downloads. All the publishers in this class advertise their URL using the textbox in the content web page and only few of them use (in addition) some of the other techniques.
Furthermore, most of them (70\%) publish only porn content.
Specifically, all those running a hosting images portal publish
exclusively porn content. By closely inspecting these sites we realize
that they store adult/sexual pictures, hence by publishing porn
content in major BitTorrent portals they are targeting millions of
users that may be interested in their web sites.
The income of the portals within this class is based on advertisement.

\noindent 
\textbf{Altruistic Publishers}: The remaining top publishers (52\%) appear to be altruistic
users since they do not seem to directly promote any URL. They are responsible of 11.5\% of the content and also 11.5\% of the downloads.
Many of these users publish music and e-book files that are light files that require lower seeding resources. Furthermore, they typically include a very extensive description of the content and often ask other users to help with seeding the content. 
Evidence suggests therefore that these publishers may have limited resources and thus they need  the help of others to sustain the distribution of their content.

\vspace{0.2cm}

In summary, half of the top publishers advertise a web portal in their
published torrents. Thus it seems that their intention, by massively
content publication via major BitTorrent portals, is to attract a large number of users to their web sites. The income of these portals come from ads and in the specific case of private BitTorrent portals also from donations and VIP accesses. Overall, these profit-driven publishers publish 26\% of the content and receives 40\% of the downloads. Therefore, the removal of these few publishers would have a dramatic impact on the demographics of the current BitTorrent open ecosystem.
In addition, there are a few altruistic publishers that sustain a relevant (but not dominant) portion of content as well as the downloads (11.5\%). This suggests that there are some \emph{good citizens} that dedicate their resources to share content with a large number of peers in spite of the legal implications this activity may have.

\subsection{Longitudinal View of Major Publishers}
So far we focused on the contributions of major publishers only during our measurement
intervals. Having identified the top publishers  in our \emph{pb10} dataset, we 
examine the longitudinal view of the contribution by major publishers since they
appear on The Pirate Bay portal. Toward this end, for each top publisher, 
we obtain the username page on The Pirate Bay portal that maintains the information
about all the published content and its published time by the corresponding user
till our measurement date (June 4, 2010)\footnote{Note that we cannot collect information about fake publishers since the web pages of their associated usernames are removed by The Pirate Bay just after identifying they are publishing fake content.}. 
Using this information for all top publishers, we capture their publishing
pattern over time with the following parameters:
{\em (i) Publisher Lifetime} which represents the number of days between the first and the last appearance of the publisher in The Pirate Bay portal, 
{\em (ii) Average Publishing Rate} that indicates the average number of published content
per day during their life time.

\begin{table}
\small
\centering
\begin{tabular}[htb]{l||l|l}
               &  Lifetime      & Avg. Publishing Rate  \\ 
	           &  (days)	    &(contents per day)     \\\hline\hline
BT Portals     & 63/466/1816    & 0.57/11.43/79.91       \\
Other Web sites & 50/459/1989    & 0.38/4.31/18.98         \\ 
Altruistic Publishers    & 10/376/1899    & 0.10/3.80/23.67          \\ 
\end{tabular}
\caption{Lifetime and Avg. Publishing Rate for the different classes of content publishers: BitTorrent Portals, Other Web sites and Altruistic Publishers. The represented values are min/avg/max per class.}
\vspace{-0.5cm}
\label{tab:longitudinal}
\end{table}

Table \ref{tab:longitudinal} shows the min/avg/max value of these metrics for the different class of publishers: BitTorrent Portals, Other Web Sites and Altruistic publishers.
The profit-driven publishers (\ie~ BitTorrent Portals and Other Web sites classes) have been publishing content during 15 months in average (at the time of the measurement) and the most longed-lived ones keep feeding content since more than 5 years ago. Furthermore, these publishers have a high publishing rate, that is surprisingly high in the case of the BT Portals class with users publishing up to 80 contents per day.
The altruistic publishers present a shorter lifetime and a lower publishing rate that seems to be due to the less motivating incentives and the lower resources they have.

In short, the content publishing in BitTorrent seems to be a profitable business since quite a long time given the lifetime of BitTorrent itself. Furthermore, the heavy seeding activity performed by profit-driven publishers (\eg~BitTorrent Portals class) during a long period of time incurs a high and continuous investment in resources that should be (at least) covered by the income from ads (and other described means) of their web portals. We analyze the income of the profit-driven publishers in the next subsection.

%

\subsection{Estimating Publishers' Income}
Our evidence obtained in previous subsections suggests that half of the top
publishers seed content on major BitTorrent portals in order to attract downloaders
to their own web site. We also showed that most of these publishers seem to generate
income at least by posting ad in their web site. In essence these publishers have 
clear financial incentives to attract users despite the cost and legal implications. 
In order to validate this key point, we assess their ability to generate income by
estimating three important but related properties of their promoting web sites:
{\em (i)} average value of the web site,
{\em (ii)} average daily income of the web site,
and {\em (iii)} average daily visits to the web site.
We obtain this information from several web sites that monitor and report these statistics
for major web sites on the Web.
To reduce any potential error in the provided statistics by individual monitoring web sites,
for each publisher's website we collect this information from six independent monitoring web site and use the
average value of these statistics across these webs\footnote{\url{www.sitelogr.com}, \url{www.cwire.com}, \url{www.websiteoutlook.com}, \url{www.sitevaluecalculator.com}, \url{www.mywebsiteworth.com},
\url{www.yourwebsitevalue.com}.}.


Table \ref{tab:economic} present the min/median/avg/max value of the previous described metrics for each one of profit-driven publisher classes: BitTorrent Portals and Other Web sites. Considering the median values (a more robust metric given the extreme values of the min and max samples) we can state that BitTorrent publishers' web sites are fairly profitable: valued in few tens thousands dollars with daily incomes of few hundred dollars and few tens thousands of visits per day. Furthermore, few publishers ($<$10) are associated to very profitable web sites valued in hundred of thousand or even millions of dollars, that receive daily incomes of thousands of dollars and hundreds of thousands visits per day.


\begin{table}

\scriptsize
{\hspace{-0.5cm}
\begin{tabular}[htb]{l||l|l|l}
               & Web site          & Website      & Website   \\ 
	           & Value (\$)       & Daily Income (\$)  & Daily visits   \\\hline\hline
BT	           &                  &               &                  \\
Portals     & 1K/33K/313K/2.8M & 1/55/440/3.7K & 74/21k/174k/1.4M \\
Other        &                  &               &                  \\
Webs        & 24/22K/142K/1.8M & 1/51/205/1.9K & 7/22K/73.5K/772K  \\ 

\end{tabular}
}
\caption{Publisher's website value (\$), daily income (\$) and num of daily visits for the different classes of profit-driven content publishers: BitTorrent Portals and Other Web sites. The represented values are min/median/avg/max per class.}
\vspace{-0.5cm}
\label{tab:economic}
\end{table}

%% file: bussiness_model.tex
\section{Other beneficiaries in BitTorrent Marketplace?}
\label{sec:bussiness}
In previous sections we analyzed the main characteristics of major
content publishers in BitTorrent, demonstrating that content
publishing is profitable \emph{business} for an important fraction of
the top publishers; a business that is responsible of 40\% of the
downloads. However, although content publishers are the key piece,
there are also other players who help sustain the business and obtain financial benefits: \emph{Major BitTorrent  Portals}, \emph{Hosting Providers} and \emph{ads companies}. Next we briefly describe their role in BitTorrent content publishing picture:

\noindent \textbf{-\emph{Major Public BitTorrent Portals}} such as The Pirate Bay are dedicated to index torrent files. They are rendezvous points where content publishers and clients publish and retrieve torrent files respectively. The main advantage of these  major portals is that they offer a reliable service (\eg~they rapidly reacts to remove fake or infected content). All this makes that millions of BitTorrent users utilize these portals every day.  These portals are the perfect target for profit-driven publisher in order to publish their torrents and advertise their web sites (potentially) to millions of users. Therefore, these major portals are one of the key players of the BitTorrent Ecosystem \cite{ross_bt} that brings substantial financial profit. For instance, The Pirate Bay is one of the most popular sites in the whole Internet (ranked the 99$^{th}$ in the Alexa Ranking) as well as one of the most valued ones (around 10M\$).

\noindent \textbf{-\emph{Hosting Providers}} are companies dedicated
to renting servers. Heavy seeding activity performed by some
publishers requires significant resources (\eg~bandwidth and storage).
Thus a large fraction of major publishers obtain the needed resources
from rented servers in hosting providers who receive income in return
for the offered service. Let's focus on OVH, the ISP responsible of a
major portion of content published in BitTorrent. Our measurement
study shows that OVH contributes between 78 and 164 different servers
(\ie~IP addresses) across the different datasets. Considering the cost
of the average server offered by OVH in its web page (around 300
euros/month) we estimate the average income obtained by OVH due to
BitTorrent content publishing ranges between roughly 23.4K and 42.9K euros/month. 
It is worth noting that some hosting providers have defined strict
policies against sharing copyrighted material through P2P applications
using their servers due to legal
issues\footnote{\url{http://www.serverintellect.com/terms/aup.aspx}.}.
However the income obtained by some hosting providers such as OVH
seems to justify the risk of potential legal actions taken against them.

\noindent \textbf{-\emph{Ads companies}} are responsible for
advertisements in the Internet. They have a set of custumers who wish
to be advertised in the Internet and a set of web sites where they put
their customers ads. They apply complex algorithms to select  where
(in which web site) and when to put each ad dynamically. They charge
their custumers for this service and part of this income is forwarded
to the web sites where the ads have been posted. Therefore, ad
companies look for popular web sites for where to put ads for their
costumers. We have demonstrated in this paper that profit-driven
BitTorrent content publisher's web sites are popular, thus most of
them post ads from ads companies\footnote{We have validated this by
looking at the header exchange between the browser and the publishers'
web site servers.}. Hence, part of the income of ads companies is
directly linked to the BitTorrent content publishing. Unfortunately,
there is no practical way to estimate the value of this incomes.

\begin{figure}[tb]
\centering
\includegraphics[height = 2in, width=3.3in]{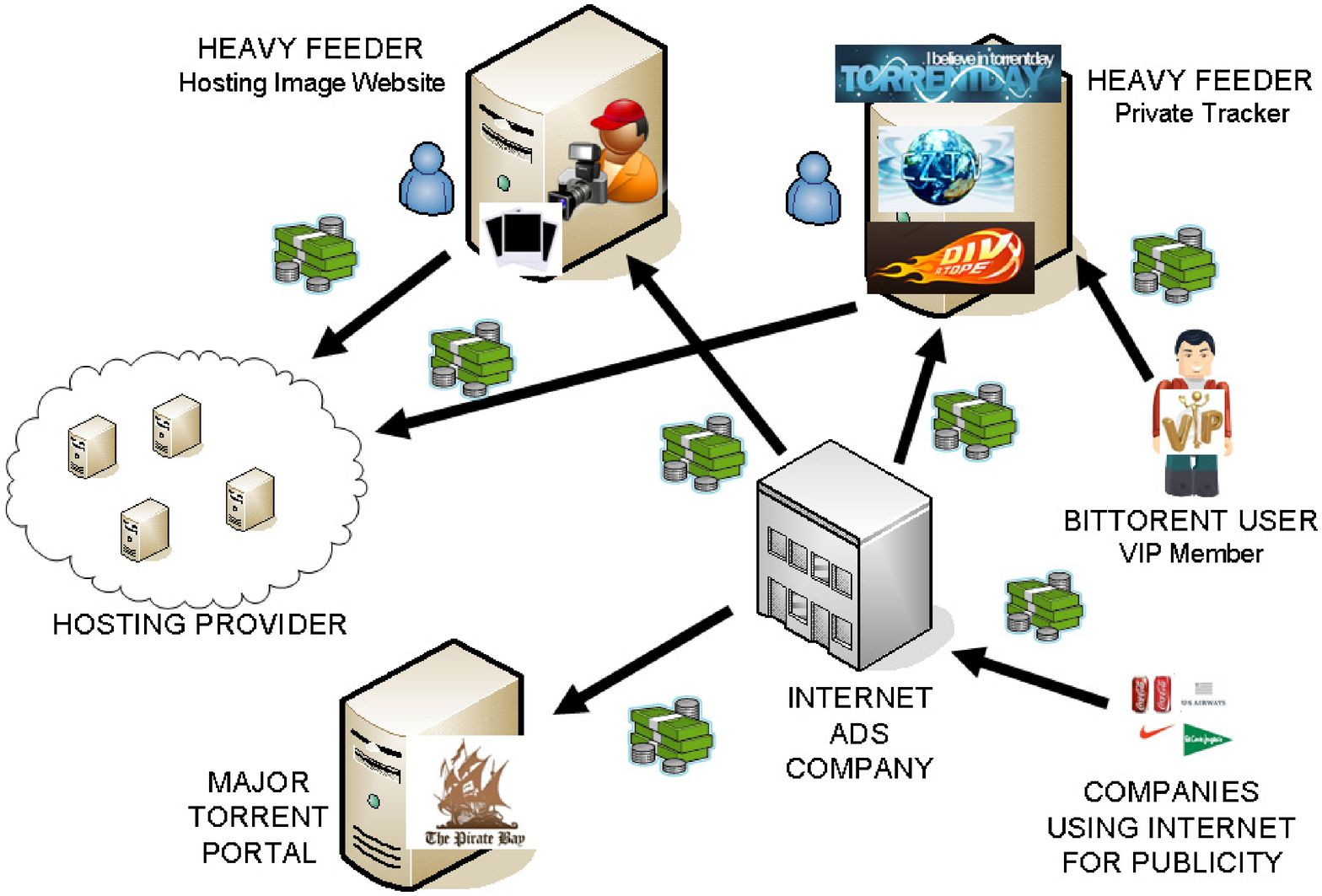}
\caption{Business Model of Content Publishing in Bittorrent}
\label{fig:business_model}
\vspace{-0.2cm}
\end{figure}

In a nutshell, this section describes the complete business model
behind content publishing in BitTorrent and briefly characterizes the main players. Finally, Figure \ref{fig:business_model} graphically represents the business model of content publishing in BitTorrent where the arrows indicate the flow of money between  the different players.

%% file: application.tex
\section{Software for Content Publishing Monitoring}
\label{sec:application}

We have implemented a  system that continuously monitors new contents
published via The Pirate Bay portal in order to retrieve information
about content publishers in real time. Our methodology is described in
Section \ref{sec:method}. Here we focus on tracking the content
publisher (and not the downloaders), thus we make only one connection
to the tracker just after we learn of a new torrent from The Pirate
Bay RSS feed. Our system retrieves the following information about
each item: filename, content category and subcategory (based on The Pirate Bay categories), publisher's username, and (in those cases we can) the publisher's IP address as well as the ISP, City and Country associated to this IP address. Furthermore, for profit-driven publishers described in this paper we have created an individual publisher's web page that provides specific information such as the publisher's promoted url or business type. The system stores all this information in a database. Finally, we have built a simple web-based interface to query this database. This interface is publicly available\footnote{ \small{\url{http://bittorrentcontentpublishers.netcom.it.uc3m.es/}}. Please, contact the authors in order to get an user and a password to log in.}.

Our application has a double goal. One the one hand, we want to share
this data with the research community to permit further analysis of
different aspects of the BitTorrent content publishing system. On the
other hand, we believe that this application can be useful for regular
BitTorrent clients. First, a BitTorrent client can easily identify
those publishers that publish content aligned with her interest (\eg~
an e-books consumer could find publishers responsible for publishing
large numbers of e-books). Furthermore, we are working on implementing
a feature to filter out fake publishers, allowing BitTorrent users of
our application (in the future) to avoid downloading fake content.


%% file: related_work.tex
\section{Related Work}
\label{sec:rw}

Significant research effort has been dedicated to understand different aspects of BitTorrent by gathering data from live swarms \cite{biersackbittorrent,Pouwelse05:BitTorrent, kaune-2009, Guo05:BitTorrentMeasurement, Menasche09:bundling, ross_bt, legout:spying, Cuevas2009:LocalityTech}. Most of these works have been dedicated to understand different demographics \cite{biersackbittorrent,Pouwelse05:BitTorrent,ross_bt} and technical \cite{Menasche09:bundling,kaune-2009,Cuevas2009:LocalityTech} aspects. However, to the best of the authors knowledge the study of socio-economics aspects of BitTorrent  (such us those covered in this paper) has received little attention.
The most relevant work to this paper is the recent study performed by LeBlond et al. \cite{legout:spying}. The authors aim to demonstrate the weakness of BitTorrent privacy. As part of their study the authors look at the demographics of BitTorrent content publishing concluding that it presents a highly skewed distribution and the important presence of hosting providers in this activity. This validates some of our initial observations. 
It is also worth mentioning the work done by Chao et al. \cite{ross_bt}. The authors present the most extensive study to characterize the BitTorrent ecosystem performed so far. A small part of the paper is also dedicated to the analysis of content publishing demographics. The authors assume that the content publisher is identified by the username that publish the torrent file in the BitTorrent portal. We have shown that this assumption may hide some important information (\eg~fake publishers). The study also corroborates that content publishing follows a skewed distribution.
Our work goes beyond simply studying the demographics of content publishing. We identify, characterize and classify the major publishers' and (more interestingly) unreveal their incentives and the business model behind content publishing in BitTorrent. 

%% file: conclusion.tex
\section{Conclusion}
\label{sec:conclusion}

In this paper we have deeply studied the content publishing activity
in BitTorrent both from a technical and a socio-economic point of
view. The results reveal that just few publishers (around 100) are
responsible for 2/3 of the published content and 3/4 of the downloads
in our dataset. We have carefully examined these users and discovered that: 
\bi
 \item antipiracy-agencies and malicious users perform systematic
poisoning index attacks over major BitTorrent Portals in order to
obstruct download of copyrighted content and to spread malware respectively. Overall, this attack contributes 30\% of the content and attracts 25\% (several millions) of downloads. 

\item BitTorrent owes an important part of its success, 1/3 of the
content and 1/2 of the downloads (if we do not consider fake content),
to a few publishers that are motivated by financial incentives.  We
believe that (sooner or latter) these few publishers will be targeted
by legal actions, and it is unclear how this will impact the global
BitTorrent ecosystem: \emph{will these major publishers stop content
publishing activity?}, in this case, \emph{will BitTorrent survive as the most important file-sharing application without these publishers?}.

\ei

%% file: acknowledgement.tex
\section{Acknowledgement}

The authors would like to thank Jon Crowcroft for his insightful comments and David Rubio for his help on developing the web-based application.

%% file: seeding_time_estimation.tex
\section{Estimation of session duration}
\label{sec:seeding_time}

In this appendix we explain the procedure utilized to calculate the duration of the session time of a given peer in a given torrent. We explain the procedure using the \emph{mn08} dataset. Note that it would be similar for \emph{pb10}.

Our \emph{mn08} crawler connects to the tracker periodically and obtains a random subset of all the IP addresses participating in the torrent. Then, we cannot guarantee to obtain the IP address of the target peer in a resolution of seconds or even few minutes. This imposes some restrictions to compute the content publisher's seeding time in a given torrent.

Therefore, we firstly define a model to estimate the number of queries to the tracker ($m$) needed to obtain the IP address of the content publisher with a given probability $\cal P$.  Let's assume that: $(i)$ we have a torrent with $\cal N$ peers and $(ii)$ for each query the tracker gives us a random set of $\cal W$ IP addresses. Then, if the target peer is in the torrent, the probability ($\cal P$) of obtaining its IP address in $m$ consecutive queries to the tracker is given by:

\begin{equation}
 {\cal P} = 1 - \left(1- \frac{{\cal W}}{{\cal N}}\right)^m
 \label{eq:session_time}
\end{equation}

We have computed the maximum instantaneous population of the torrents in our \emph{mn08} and found that 90\% of torrents have tyically less than 165 concurrent peers. Then, we assume that the torrents have always a population  $\cal N$ = 165. This is an upper bound that allows us to remove the noise introduced by the churn. We make a second conservative assumption: the tracker gives us $\cal W$ = 50 random IPs in each response (although in most of the cases we obtain 200 IP addresses). With these numbers and the proposed model we can assure that, if a peer (\eg~ a content publisher) is in the torrent, we will discover it in $m$ = 13 queries to the tracker with a probability higher than 0.99.

Next, we have calculated the time between 2 consecutive queries to the tracker in our dataset, and have checked that 90\% of them are less than 18 minutes apart. Then, we again make a conservative assumption and consider that the time between two consecutive queries is 18 minutes.

Hence, multiplying the number of needed queries by the time between two consecutive queries we conclude that if a peer (\eg~content publisher) is in the torrent, we are able to get  its IP address in a period of 4h with a probability  equal to 0.99. Therefore, we consider that a given content publisher is offline (\ie~its session has finished) if its IP address is not gathered in the torrent during 4 hours. We have repeated the experiments with 2h and 6h thresholds obtaining similar results.